\documentstyle[12pt]{article}

\newcommand{\be}{\begin{equation}}
\newcommand{\ee}{\end{equation}}
\newcommand{\ba}{\begin{eqnarray}}
\newcommand{\ea}{\end{eqnarray}}
\newcommand{\bas}{\begin{eqnarray*}}
\newcommand{\eas}{\end{eqnarray*}}

\newcommand{\nn}{\nonumber}

\newcommand{\bomega}{\mbox{\boldmath $\omega$}}

\newcommand{\tr}{\mbox{tr}}

\newcommand{\oh}{\frac{1}{2}}

\newcommand{\cA}{{\cal A}}
\newcommand{\cB}{{\cal B}}
\newcommand{\cC}{{\cal C}}
\newcommand{\cD}{{\cal D}}

\begin{document}
 
\addtolength{\baselineskip}{0.20\baselineskip}
\hfill UT-Komaba 98-5
 
\hfill Feburary 1998
\begin{center}
 
\vspace{36pt}
{\large \bf  Determinant formula for  \\
the six-vertex model with reflecting end}
 
\end{center}

\vspace{36pt}
 
\begin{center}

\vspace{5pt}
 Osamu Tsuchiya \footnote{E-mail address: otutiya@hep1.c.u-tokyo.ac.jp} \\
\vspace{3pt}
{\sl Institute of Physics, 
University of Tokyo, \\
Komaba, Meguro-ku, Tokyo 153, Japan}

\end{center}

\vspace{36pt}
 
%\vfill
 
\begin{center}
{\bf Abstract}
\end{center}
 
\vspace{12pt}
 
\noindent
Using the Quantum Inverse Scattering Method for
the XXZ model with open boundary conditions,
we obtained the determinant formula for 
the six vertex model with reflecting end.
 
\vspace{24pt}
 
\vfill
 
\newpage
%%%%%%%%%%%%%%%%%%%%%%%%%%%%%%%%%%
 
% Introduction

\section{Introducntion}
%
%
%\noindent
%{\it Introduction} 

Determinant representations of 
correlation functions of  one dimensional quantum 
integrable models were 
studied extensively\cite{KBI}.
Especially Essler et. al. studied the correlation functions of 
XXZ model with periodic boundary conditions \cite{EFIK}.
In the calculation of scalar products of the XXZ model,
the relation between the scalar products of the model
and the partition function of the six-vertex with domain wall
boundary conditions is essential \cite{K82}.
Izergin et. al. used this relation to obtain the 
determinant formula for the partition function of 
the six-vertex model with domain wall boundary 
conditions \cite{ICK}.

On the other hand, integrable models with open boundary
conditions are interesting.
Integrable model with open boundary conditions are
related to 
the $B_{N}$ type Weyl group \cite{YT}
and in condenced matter physics they are
related to the impurity problems in the Luttinger liquid \cite{TY}.

Correlation function of the 1 + 1 dimensional
delta function interacting Bose gas
with open boundary conditions were studied in \cite{K97}.

For the XXZ model with open boundary conditions,
the algebraic Bethe ansatz of the model
can be used to calculate the partition function of 
the six-vertex model with 
reflecting boundary condition at one boundary 
and domain wall boundary conditions
at other three boundaries.
(In this paper we call
six vertex model with this type of boundary conditions
{\it six vertex model with reflecting end}.) 
In this paper, we obtain the determinant formula
for the above six-vertex model using above relation.

%%%%%%%%%%%%%%%%%%%%%%%%%%%%%%%%%%%%%%%%%%%%%%%%%%%%%
%%%%%%%%%%%%%%%%%%%%%%%%%%%%%%%%%%%%%%%%%%%%%%%%%%%%
\section{Quantum Inverse Scattering Method for the
          Open XXZ Model}

%
%
%\noindent
%{\it Quantum Inverse Scattering Method for the Open XXZ Model}

Let us start with a brief review of the Quantum Inverse Scattering Method
for the XXZ model with open boundary conditions \cite{S}.
The trigonometrical solution of the Yang-Baxter equation
is given by
\ba
R_{0 0'} (\lambda ) 
 & = & \oh \left\{ \left[ 
\frac{\sinh (\lambda -2i\eta) -\sinh (2i\eta)}{\sinh (\lambda)} \right] 
1_{0} \otimes 1_{0'}
 + \sigma^{x}_{0} \otimes \sigma^{x}_{0'} 
 + \sigma^{y}_{0} \otimes \sigma^{y}_{0'} \right. \\
 & & \left. + \left[
\frac{\sinh (\lambda -2i\eta) + \sinh (2i\eta)}{ \sinh (\lambda) } \right]
 \sigma^{z}_{0} \otimes \sigma^{z}_{0'} \right\} ,
\ea
where $R_{0 0'}(\mu)$ acts as a linear operator on the
tensor product of two auxiliary spaces.
The $L$ operators are given by 
$ L_{n} (\lambda ) = \sinh (\lambda) R_{0n}(\lambda) P_{0n}$
where $0$ denotes the auxiliary space, $n$ denote the $n$-th
quantum space and 
\newpage
$P_{0n} = (1/2) (1_{0} \otimes 1_{n}
 + \sigma^{x}_{0} \otimes \sigma^{x}_{n} 
 + \sigma^{y}_{0} \otimes \sigma^{y}_{n} 
+ \sigma^{z}_{0} \otimes \sigma^{z}_{n})$ is a  permutation operator.
Thus
\ba
L_{n} (\lambda ) 
 & = & \oh \left\{ \left[ 
\sinh (\lambda -2i\eta) +\sinh (\lambda) \right] 
1_{0} \otimes 1_{n}
 - \sinh(2i\eta) \left[ \sigma^{x}_{0} \otimes \sigma^{x}_{n} 
 + \sigma^{y}_{0} \otimes \sigma^{y}_{n} \right] \right. \nn \\
 & & + \left. \left[
\sinh (\lambda -2i\eta) - \sinh (\lambda) \right]
 \sigma^{z}_{0} \otimes \sigma^{z}_{n} \right\}.
\ea

As a consequence of the Yang-Baxter equation,
the $L$ operators satisfy the fundamental commutation relations
\ba
R (\lambda -\mu )( L_{n}(\lambda) \otimes L_{n}( \mu)) 
 & = & ( L_{n}(\mu) \otimes L_{n}( \lambda)) R (\lambda -\mu ).
\label{fcr}
\ea

The monodromy matrix of 
inhomogeneous XXZ model is constructed from  the $L$ operators as:
\ba
T(\lambda ,  \bomega) & = &
L_{L} (\lambda -\omega_L ) L_{L-1} (\lambda -\omega_{L-1}) 
\cdots L_{1} (\lambda -\omega_{1}).
\ea
This matrix  satisfies  fundamental commutation relations similar to 
(\ref{fcr})
\ba
R (\lambda -\mu )( T(\lambda, \bomega) \otimes T ( \mu, \bomega)) 
 & = & ( T (\mu,\bomega) \otimes T (\lambda,\bomega) R (\lambda -\mu ).
\ea

For the model with open boundary conditions,
in addition to the Yang-Baxter equation, the reflection equation \cite{S}
plays an important role,
\ba
R_{12} (\lambda - \mu) K_{1}(\lambda) R_{12}(\lambda +\mu)
 K_{1} (\mu)
& = &
 K_{1} (\mu) R_{12}(\lambda +\mu)
K_{1}(\lambda)R_{12} (\lambda - \mu),
\ea
where $K_{j}(\lambda)$ is the $2\times2$ matrix
which act to the $j$-th auxiliary space and
is related to the reflection of the particle 
with rapidity $\lambda$
at the boundaries. 
In the present paper, we concentrate to the diagonal solution
of the reflection equation
\ba
K (\lambda, \xi) & = & 
                         \frac{1}{\sinh (\xi)} 
                        \left( \begin{array}{cc}
                               \sinh (\xi + \lambda) & 0 \\
                                0 & \sinh (\xi -\lambda)
                              \end{array}   \right).
\ea

For the model with open boundary conditions,
Sklyanin`s monodromy matrix
\be
U (\lambda, \bomega)  =  T (\lambda, \bomega ) 
K (\lambda , \xi_{-}) \tilde{T}(\lambda, \bomega) 
= \left( \begin{array}{cc}
          \cA (\lambda, \bomega) & \cB (\lambda,\bomega) \\
          \cC (\lambda,\bomega) & \cD (\lambda,\bomega)
          \end{array}       \right),
\ee
where
\ba
&&
\tilde{T}(\lambda, \bomega)
=
L_1 (\lambda +\omega_1) L_2 (\lambda +\omega_2) 
\cdots L_L (\lambda+\omega_L) \nn \\
&&
\propto T^{-1} (-\lambda,\bomega)
\ea
 is fundamental,
where $ \cA (\lambda,\bomega), \cB (\lambda,\bomega), 
\cC (\lambda,\bomega), \cD (\lambda,\bomega)$
are operators which act on the quantum spaces.

$U(\lambda,\bomega)$ also satisfies reflection equation
\ba
&&
R_{12} (\lambda - \mu) U_{1}(\lambda,\bomega) 
R_{12}(\lambda +\mu) U_{1} (\mu,\bomega) \nn \\
&&
= 
 U_{1} (\mu,\bomega) R_{12}(\lambda +\mu)
U_{1}(\lambda,\bomega)R_{12} (\lambda - \mu).
\ea

The reflection equation is rewritten as the commutation relations
of the matrix elements $\cA, \cB, \cC, \cD$,
\ba
%%%%%%%%%% [A,A],[A,D],[D,D]
%
 && [ \cA (\lambda) , \cA (\mu) ]  =  - [\cD (\lambda),\cD (\mu)] \nn \\
 && = \frac{\sinh (2i\eta)}{\sinh (\lambda +\mu -2i\eta)}
   \{\cB (\lambda) \cC (\mu) -\cB (\mu) \cC (\lambda) \}\nn \\
&& = -  \frac{\sinh (2i\eta)}{\sinh (\lambda +\mu -2i\eta)}
   \{\cC (\lambda) \cB (\mu) -\cC (\mu) \cB (\lambda) \}\nn \\
&& = \frac{\sinh (\lambda - \mu) \sinh (\lambda +\mu)}{
             \sinh^{2}(2i \eta) -\sinh (\lambda +\mu -2i \eta)}
            [\cA (\lambda) , \cD (\mu) ]   \\
\label{comad}
%
%
%%%%%%%%%%%% [B,B],[C,C]
%
& & [\cB ( \lambda), \cB ( \mu) ] = [\cC (\lambda), \cC (\mu)] = 0 \\
\label{combc1}
%
%
%
%%%%%%%%%%%% [B,C]
&&
[\cB (\lambda) , \cC (\mu)] = \frac{\sinh (2i\eta)}{
\sinh (\lambda -\mu) \sinh (\lambda +\mu -2i\eta)}\times  \nn \\
&&
 \times \{ 
\sinh (\lambda +\mu) [ \cD (\lambda)\cA (\mu) 
- \cD (\mu) \cA (\lambda) ] \nn \\
&&
~~~~~~~~~ + \sinh (\lambda -\mu)[ \cA (\lambda)\cA (\mu) 
- \cD (\mu) \cD (\lambda) ]
\}         \\
\label{combc2}
\ea
%
%
%%%%%%%%%%[A,B]
%&&
%\cA(\mu) \cB (\lambda) =
% \frac{\sinh(\lambda -\mu -2i\eta)\sinh(\lambda+\mu)
%}{\sinh (\lambda +\mu -2i\eta)\sinh(\lambda -\mu)} \cB(\lambda)\cA(\mu) \nn \\
%&&
%+  \frac{\sinh(2i\eta)\sinh(\lambda+\mu)
%}{\sinh (\lambda +\mu -2i\eta)\sinh(\lambda -\mu)} \cB(\mu) \cA(\lambda) \nn \\
%&&
%+ \frac{\sinh(2i\eta)
%}{\sinh (\lambda +\mu -2i\eta)} \cB (\mu) \cD (\lambda) \\
%\label{comab}
%%%%%%%%%%%%%%%%%%%%%%%%% [D,B]
%&&
%\cD (\lambda) \cB (\mu) =
%-2 \frac{\sinh^{2} (2i\eta) \cosh (2i\eta)}{
%\sinh (\lambda -\mu) \sinh (\lambda + \mu -2i\eta)} 
%\cB (\mu) \cA (\lambda) \nn \\
%&&
%- \frac{\sinh (2i\eta) \sinh (\lambda - \mu -4i\eta )}{
%\sinh (\lambda -\mu) \sinh (\lambda + \mu -2i\eta)} 
%\cB(\lambda) \cA (\mu) \nn \\
%&&
%+ \frac{\sinh (2i\eta) \sinh (\lambda + \mu -4i\eta )}{
%\sinh (\lambda -\mu) \sinh (\lambda + \mu -2i\eta)} 
%\cB (\lambda) \cD(\mu) \nn \\
%&&
%+ \frac{\sinh (\lambda - \mu -2i\eta ) \sinh (\lambda + \mu -4i\eta )}{
%\sinh (\lambda -\mu) \sinh (\lambda + \mu -2i\eta)} 
%\cB(\mu) \cD(\lambda) 
%\label{comdb}
%\ea

From the above commutation relations (\ref{comad})
Sklyanin`s transfer matrices
\ba
&&
\tau ( \lambda)= \tr_{0} (K(-\lambda+ 2i\eta) U( \lambda)) \\
&&
= (1/\sinh(\xi_{+})) [ \sinh (\xi_{+} -\lambda +2i\eta) \cA (\lambda)
                      + \sinh (\xi_{+} +\lambda - 2i \eta) \cD (\lambda) ]
\ea
are commutative for any values of spectral 
parameter $[\tau (\lambda),\tau (\mu)]$.

Consequently, $\tau (\lambda)$ generate the infinite number of
 conserved quantities.
For example the Hamiltonian is given by
\ba
&&
\frac{d}{d\lambda} \tau (\lambda) |_{\lambda =0}  \nn \\
&&
= - \frac{1}{\sinh (2i\eta)} \sum_{n=1}^{L-1} 
\{  \sigma_{n}^{x} \sigma_{n+1}^{x}
+ \sigma_{n}^{y} \sigma_{n+1}^{y}
+ \cosh (2i\eta) \sigma_{n}^{z} \sigma_{n+1}^{z}  \} \nn \\
&&
+\coth(\xi_{-})\sigma_{1}^{z} -\coth(\xi_{+}) \sigma_{L}^{z} 
+\mbox{constant}
\ea

In the algebraic Bethe ansatz,
we construct the simultaneous eigenstates of the
conserved quantities by applying the some elements (creation operator)
of the monodromy matrix to the pseudovacuum.
We choose as a pseudovacuum the state whose 
spins are all up
\ba
&&
|0 \rangle = \otimes_{j=1}^{L} | + \rangle_{j}.
\ea
This pseudovacuum is the eigenstate of 
$\cA$ and $\cD$ and is annihilated by $\cC$
\ba
&&
\cA (\lambda) |0\rangle =  a (\lambda)|0\rangle, \nn \\
&&
\cD (\lambda) |0\rangle =  d (\lambda)|0\rangle, \nn \\
&&
\cC (\lambda) |0\rangle = 0,
\ea
where
\ba
&&
 a(\lambda) = \sinh (\xi_{-}+ \lambda) \prod_{n=1}^{L}
\{ \sinh (\lambda - \omega_{n} -2i\eta) \sinh (\lambda +\omega_{n}-2i\eta) \},
\nn \\
&&
 d (\lambda) = \sinh (\xi_{-}- \lambda) \prod_{n=1}^{L}
\{ \sinh (\lambda - \omega_{n}) \sinh (\lambda +\omega_{n}) \}.
\ea
In other words $\cB$ has the role of  the creation operator 
and the eigenstates of the Hamiltonian are of the form 
$ \prod_{n=1}^{N} \cB (\lambda_{n}, \bomega) |0 \rangle $
if rapidities $\lambda_{n}$ satisfy the
Bethe ansatz equations.

%%%%%%%%%%%%%%%%%%%%%%%%%%%%%%%%%%%%%%%%%%%%%%%%%%%%%%%%%%%
In the algebraic Bethe ansatz 
eigenstates and dual states of eigenstates  are respectively 
given by
\ba 
&&
\cB(\lambda_{L})\cdots \cB(\lambda_{1}) |0 \rangle  
\ea
and 
\ba
&&
\langle 0 |\cC (\mu_{1}) \cdots \cC (\mu_{L}) .
\ea

%%%%%%%%%%%%%%%%%%%%%%%%%%%%%%%%%%%%%%%%%%%%%%%%%%%%%%%%%%%%%%%%%%
%%%%%%%%%%%%%%%%%%%%%%%%%%%%%%%%%%%%%%%%%%%%%%%%%%%%%%%%%%%%%%%%%%%
%
\section{Inhomogeneous six-vertex model with reflecting end}

%\noindent
%{\it Inhomogeneous six-vertex model with reflecting end}

For the scalar products between the eigenstates
the following relation holds
\ba
&&
\langle 0 |\cC (\mu_{1}) \cdots \cC (\mu_{L}) 
\cB(\lambda_{L})\cdots \cB(\lambda_{1}) |0 \rangle \nn \\
&&
= \langle 0 |\cC (\mu_{1}) \cdots \cC (\mu_{L}) | \bar{0} \rangle
\langle \bar{0} | \cB(\lambda_{L})\cdots \cB(\lambda_{1}) |0 \rangle
\ea
where $ |\bar{0} \rangle $ is the quantum state
with all spins are down $\prod_{n=1}^{L} | - \rangle$.
Note that 
\ba
&&
Z_{L} ( \{\lambda \}, \{\omega\})
=  \langle \bar{0} | \cB(\lambda_{L})\cdots \cB(\lambda_{1}) |0 \rangle,
\ea
is nothing but  the partition function
of the inhomogeneous 6 vertex model
on a $L \times 2L $ lattice
with domain wall boundary condition for
the lower, upper and left boundaries and
 reflecting end for the right boundary
(see Fig. 1.).

$\cB(\lambda_{n}) 
=  _{n}^{0} ( +| U(\lambda,\omega)|-)^{0}_{n}$
where $|+)^{0}_{n}$ and $|-)^{0}_{n}$ 

are the $n$-th auxiliary states with spins
$+$ and respectively $-$
($^{0}_{n}(+|$ and $^{0}_{n}(-|$ are dual states 
with spins $+$ and respectively $-$).
Here we assign the auxiliary space to each $\cB (\lambda_{n})$.

%%%%%%%%%%%%%%%% recursion relation
%
Now we derive the recursion relations to the partition function
$Z_{L}(\{\lambda\},\{\omega\})$.

If the inhomogenuity $\omega_{L}$ is choosen to be
$\omega_{L} = -\lambda_{1}$ 
then the partition function of the $L \times 2L$ lattice
is related to the partiton function of the
$L-1 \times 2(L-1)$ lattice.

The $L$ operators are regular i.e.
\ba
&&
L_{L} (0) = - \sinh (2i\eta) P_{oL}.
\ea

Thus 
\ba
&&
\cB (\lambda_{1}) \prod_{j=1}^{L} |+>_{j} |_{\omega_{L}=-\lambda_1} \nn \\
&&
= -\sinh (2i\eta) {}^{o}_{1}(+|L_{L}(\lambda_{1}-\omega_{L}) \cdots
  L_{L-1} (\lambda_{1} +\omega_{L-1}) P_{oL} |-)^{o}_{1} 
\prod_{j=1}^{L} |+\rangle_{j} \nn \\
&&
= -\sinh (2i\eta) {}^{o}_{1}(+|L_{L}(\lambda_{1}-\omega_{L}) \cdots
  L_{L-1} (\lambda_{1} +\omega_{L-1})  |+)^{o}_{1} 
|-\rangle_{L} \prod_{j=1}^{L-1} |+ \rangle_{j}. \nn \\
\ea

Noting that  $ |+)_{1} |+\rangle_{n} $ is the eigenstate
of $L_{n}(\lambda_{1})$
\ba
&&
L_{n}(\lambda_{1}) |+)_{1} |+\rangle_{n} 
= \sinh (\lambda_{1} -2i\eta)|+)_{1} |+\rangle_{n},
\ea
 
\ba
&&
K(\lambda_{1} , \xi_{-}) |+)_{1} 
= \frac{\sinh (\xi_{-} +\lambda_{1})}{\sinh (\xi_{-})} |+)_{1}
\ea
and
\ba
&&
 _{1}(+| L_{L} (\lambda_{n} -\omega_{L}) |-\rangle_{L}
= 
2\sinh (\lambda_{n} -\omega_{L})  {}_{1}(+| |-\rangle_{L}
\label{aplusqminus}
\ea

We obtain
\ba
&&
\cB (\lambda_{1}) \prod_{j=1}^{L} |+>_{j} |_{\omega_{L}=-\lambda_1} \nn \\
&&
=
- \frac{\sinh(\xi_{1} +\lambda_{1})}{\sinh (\xi_{-})}
\sinh (2i\eta) \sinh (2 \lambda_{1}) \nn \\
&&
\times \prod_{j=1}^{L-1} \sinh(\lambda_{1} -\omega_{j}-2i\eta)
                \sinh(\lambda_{1} +\omega_{j}-2i\eta) \nn \\
&&
 \times |- \rangle_{L} \prod_{j=1}^{L-1} |+\rangle_{j},
\ea

Using 
\ba
&&
L_{L} (\lambda_{n} + \omega_{L}) |- )_{n} |- \rangle_{L}
= \sinh (\lambda_{n} + \omega_{L}-2i\eta ) |- )_{n} |- \rangle_{L},
\ea
together with equation (\ref{aplusqminus})
and the locality of the $L$ operators 
we find the relation
\ba
&&
\cB (\lambda_{n}) |-\rangle_{L} =
\sinh (\lambda_{n}+\omega_{L}-2i\eta ) \sinh (\lambda_{n}-\omega_{L})
|-\rangle_{L} \cB_{L-1} (\lambda_{n}),
\ea
where
\ba
\cB_{L-1}(\lambda_{n}) 
&
=
&
{}_{n}(+| L_{L-1}(\lambda_{n} -\omega_{L-1}) \cdots
L_{1}(\lambda_{n} -\omega_{1})
K (\lambda_{n}, \xi_{-}) \nn \\
&&
\times
L_{1}(\lambda_{n} + \omega_{1}) \cdots
L_{L-1}(\lambda_{n} +\omega_{L-1}) |-)_{n}.
\ea

%Finally the partition function
%of the model on the $L \times 2L $ lattice 
%have the relationship to that on the 
%$L-1 \times 2(L-1)$ lattice
%\ba
%&&
%Z_{L}(\{\lambda \},\{\omega\}) |_{\omega_{L} = -\lambda_{1}} \nn \\
%&&
%= 
%- \frac{\sinh(\xi_{1} +\lambda_{1})}{\sinh (\xi_{-})}
%\sinh (2i\eta) \sinh (2 \lambda_{1}) \nn \\
%&&
%\times \prod_{j=1}^{L-1} \sinh(\lambda_{1} -\omega_{j}-2i\eta)
%                \sinh(\lambda_{1} +\omega_{j}-2i\eta) \nn \\
%&&
%\times  \prod_{n=2}^{L}
%\sinh (\lambda_{n} +\omega_{L}-2i\eta ) \sinh (\lambda_{n} -\omega_{L})
%\prod_{j=1}^{L-1}\langle -| 
%\prod_{n=2}^{L} \cB_{L-1} (\lambda_{n},\omega)
%\prod_{j=1}^{L-1} |+ \rangle  .
%\ea

{\it lemma 1-a}

The partition function
of the model on the $L \times 2L $ lattice 
is related to hte one  on 
$L-1 \times 2(L-1)$ lattice
\ba
&&
Z_{L}(\{\lambda \},\{\omega\}) |_{\omega_{L} = -\lambda_{1}} \nn \\
&&
= 
- \frac{\sinh(\xi_{-} +\lambda_{L})}{\sinh (\xi_{-})}
\sinh (2i\eta) \sinh (2 \lambda_{L}) \nn \\
&&
\times \prod_{j=1}^{L-1} \sinh(\lambda_{L} -\omega_{j}-2i\eta)
                \sinh(\lambda_{L} +\omega_{j}-2i\eta) \nn \\
&&
\times  \prod_{n=1}^{L-1}
\sinh (\lambda_{n} +\omega_{L}-2i\eta ) \sinh (\lambda_{n} -\omega_{L})
Z_{L-1}(\{\lambda\},\{\omega\}).
\ea
Here $Z_{M}(\{\lambda\},\{\omega\})$ is the partition function
for the model on $M \times 2M $ lattice
\ba
&&
Z_{M}(\{\lambda\},\{\omega\}) =
\prod_{j=1}^{M} {}_{j}\langle -| \prod_{n=1}^{M} 
\cB_{M} (\lambda_{n},\bomega) 
\prod_{j=1}^{M} |+ \rangle_{j},
\ea
which depend on $\lambda_{1} \sim \lambda_{M}$
    and  $\omega_{1} \sim \omega_{M}$.
%Because of equation (\ref{combc1})
%$Z_{L}(\{\lambda\},\{\omega\})$ is
%symmetric in $\{\lambda\}$,
%we obtain the recursion relations

{\it lemma 1-b}

Folowing the same line of  argument as above
we have other type of  recursion relations
\newpage
\ba
&&
Z_{L}(\{\lambda \},\{\omega\}) |_{\omega_{L} = \lambda_{1}} \nn \\
&&
= 
- \frac{\sinh(\xi_{-} -\lambda_{L})}{\sinh (\xi_{-})}
\sinh (2i\eta) \sinh (2 \lambda_{L}) \nn \\
&&
\times \prod_{j=1}^{L-1} \sinh(\lambda_{L} -\omega_{j}-2i\eta)
                \sinh(\lambda_{L} +\omega_{j}-2i\eta) \nn \\
&&
\prod_{n=1}^{L-1}
\sinh (\lambda_{n} -\omega_{L}-2i\eta ) \sinh (\lambda_{n} +\omega_{L})
Z_{L-1}(\{\lambda\},\{\omega\}).
\ea

{\it Lemma 2}
\ba
&&
Z_{1}( \{\lambda \}, \{\omega\} )
= {}_{1}\langle \bar{0}| \cB (\lambda) |0 \rangle_{1}
\ea
is explicitly calculated as
\ba
&&
Z_{1}( \{\lambda \}, \{\omega\} )
= - \frac{\sinh (2i\eta)}{\sinh (\xi_{-})}
[ \sinh(\xi +\lambda) \sinh (\lambda -\omega)
 + \sinh(\xi -\lambda) \sinh (\lambda +\omega) ].
\ea

{\it Lemma 3}

$ \langle \bar{0}| \cB (\lambda_{L}) \cdots 
\cB (\lambda_{1}) | 0 \rangle $
depend on each $\lambda_{n}$ as
\ba
&&
\langle \bar{0}| \cB (\lambda_{L}) \cdots 
\cB (\lambda_{1}) | 0 \rangle  \nn \\
&&
=
e^{- 2L \lambda_{n}} P_{2L} (e^{2\lambda_{n}})
\ea
where $P_{2L}(x)$ is the polynomial of 
degree $2L$.

This  is proved along  the same line of argument
as in reference \cite{ICK}.

%%%%%%%%%%%%%%%%%%%%%%%%%%%%%%%%%%%%%%%%%%%%%%%
%%%%%%%%%%%%%%%%%%%%%%%%%%%%%%%%%%%%%%%%%%%%%%%
\section{Determinant formula for the six vertex model
     with reflecting end}
%
%\noindent
%{\it Determinant formula for the six vertex model
%     with reflecting end}
%
The lemmas in the last section determine
the partition function of the model uniquely.

\noindent
{\bf Theorem:}
The partition function of the six vertex model with 
reflecting end is given by
\ba
&&
Z_{L} (\{ \lambda \},\{\omega\}) \nn \\
&&
=
\prod_{j=1}^{L}\prod_{n=1}^{L}
\sinh(\lambda_{j}-\omega_{n}-2i\eta) \sinh (\lambda_{j}-\omega_{n})
\sinh(\lambda_{j}+\omega_{n}-2i\eta) \sinh (\lambda_{j}+\omega_{n})
\nn \\
&&
\times [ \prod_{j >k} \sinh(\lambda_{j} -\lambda_{k})
             \sinh(\lambda_{j} +\lambda_{k} -2i\eta)
\prod_{n>m} \sinh(\omega_{n}-\omega_{m})
             \sinh(\omega_{n}+\omega_{m}) ]^{-1}
\nn \\
&&
\times 
det M
\ea
where
\ba
&&
M_{jn} 
=
\frac{\sinh (2i\eta)}{
\sinh (\xi)
    \sinh(\lambda_{j}-\omega_{n}-2i\eta)
    \sinh(\lambda_{j}+\omega_{n} -2i\eta)} \nn \\
&&
 \times  \left\{ 
\frac{\sinh(\xi_{-}+\lambda_{j})}{\sinh(\lambda_{j}+\omega_{n})}
+
\frac{\sinh(\xi_{-}-\lambda_{j})}{\sinh(\lambda_{j}-\omega_{n})}
\right\}
\ea

%%%%%%%%%%%%%%%%%%%%%%%%%%%%%%%%%%%%%%%%%%%%%%%%%%%%%%%%%%
\section{Conclusion}

%\noindent
%{\it Conclusion}
In this paper the determinant formula 
for the partition function of the six-vertex model
with reflecting end is given.
It is the determinant formula corresponding to 
the Quantum Inverse Scattering Method with open boundary conditions.

In the model with open boundary conditions, 
as opposed to the model with periodic boundary conditions,
one can not determine the scalar product from the six-vertex model
with reflecting end.
To calculate the scalar products and correlation functions
for the open XXZ model is interesting.

\vspace{12pt}
 
%\noindent {\bf Acknowledgment:} 

\vspace{24pt}

%%%%%%%%%%%%%%%%%%%   References %%%%%%%%%%%%%%%%%%%%%%%%%

%%%%%%%%%%%%%%%%%% figure 1


\begin{thebibliography}{99}

%%%%%%%%%%%%%%%%%% correlation functions review

\bibitem{KBI}
V. E. Korepin, N. M. Bogoliubov and A. G. Izergin,
{\it Quantum Inverse Scattering Method and Correlation Function}.


%%%%%%%%%%%%%%%%%% correlation functions

%% calculation of norm

\bibitem{K82}
V. E. Korepin,
Commun. Math. Phys. {\bf 86} (1982) 391.

%% determinant representation for the XXZ model

\bibitem{EFIK}
F. H. L. Essler, H. Frahm, A. G. Izergin and V. E. Korepin,
Comm. Math. Phys. {\bf 174} (1995) 191.

%% integro difference equation for the XXZ model

\bibitem{EFIK}
F. H. L. Essler, H. Frahm, A. R. Its and V. E. Korepin,
{\it Integro-Difference Equation for a correlation function
of the spin-$1/2$ Heisenberg XXZ chain},
preprint (1995) cond-mat/9503142.

%% determinant formula in six vertex model

\bibitem{ICK}
A. G. Izergin, D. A. Coker and V. E. Korepin,
J. Phys. A: Math. Gen. {\bf 25} (1992) 4315.

%% B_{N} type Calogero model

\bibitem{YT}
T. Yamamoto and O. Tsuchiya,
J. Phys. A: Math.Gen. {\bf 29} (1996) 3977.

%% Hubbard model with boundary

\bibitem{TY}
O. Tsuchiya and T. Yamamoto,
J. Phys. Soc. Japan {\bf 66} (1997) 1950.




%% correlation functions in the nonlinear Schrodinger model with boundary
\bibitem{K97}
T. Kojima,
{\it Dynamical correlation functions for an impenetrable Bose gas
with open boundary conditions},
preprint (1997)


%%%%%%%%%%%%%%%%% Quantum Inverse Scattering Method for the model
%%%%%%%%%%%%%%%%% with open boundary conditions

\bibitem{S}
E. K. Sklyanin,
J. Phys. A: Math. Gen. {\bf 21} (1988) 2375.



\end{thebibliography}
\end{document}